\def\year{2023}\relax
\documentclass[letterpaper]{article} 
\usepackage{aaai22}  
\usepackage{times}  
\usepackage{helvet}  
\usepackage{courier}  
\usepackage[hyphens]{url}  
\usepackage{graphicx} 
\usepackage{natbib}  
\usepackage{caption} 
\DeclareCaptionStyle{ruled}{labelfont=normalfont,labelsep=colon,strut=off} 
\usepackage{algorithm}
\usepackage{algorithmic}
\graphicspath{./images/}
\usepackage{array}
\usepackage{multirow}
\usepackage[font=small,labelfont=bf]{caption}

\usepackage{newfloat}
\usepackage{listings}
\floatstyle{ruled}
\newfloat{listing}{tb}{lst}{}
\floatname{listing}{Listing}
\title{Online Networks of Support in Distressed Environments:\\Solidarity and Mobilization during the Russian Invasion of Ukraine}
\author {
    Jinyi Ye, 
    Nikhil Jindal, 
    Francesco Pierri, 
    Luca Luceri 
}
\affiliations {
    Information Sciences Institute, University of Southern
California, Los Angeles, USA\\
}

\copyrightyear{2023}

\begin{document}

\maketitle

\begin{abstract}
Despite their drawbacks and unintended consequences, social media networks have recently emerged as a crucial resource for individuals in distress, particularly during times of crisis. These platforms serve as a means to seek assistance and support, share reliable information, and appeal for action and solidarity. 
In this paper, we examine the online networks of support during the Russia-Ukraine conflict by analyzing four major social media networks---Twitter, Facebook, Instagram, and YouTube. Using a large dataset of 68 million posts, we explore the temporal patterns and interconnectedness between these platforms and online support websites. 
Our analysis highlights the prevalence of crowdsourcing and crowdfunding websites as the two main support platforms to mobilize resources and solicit donations, revealing their purpose and contents, and investigating different support-seeking and -receiving practices. Overall, our study underscores the potential of social media in facilitating online support in distressed environments through grassroots mobilization, contributing to the growing body of research on the positive impact of online platforms in promoting social good and protecting vulnerable populations during times of crisis and conflict.
\end{abstract}

\section{Introduction}
\noindent Over the past decade, digital platforms like social media and messaging apps have emerged as crucial players during times of crisis, particularly in situations involving social issues, conflicts, and wars. Individuals in distressed environments have leveraged these platforms to document human rights abuses and atrocities, appeal to the international community for action, and seek relief, assistance, and support  \cite{alexander2014social}. This trend has been observed in several countries, including but not limited to Syria \cite{alencar2018refugee, dekker2018smart}, Israel and Palestine \cite{wulf2013fighting}, Libya \cite{ali2013gatekeeping}, Haiti \cite{liu2014crisis}, and Myanmar \cite{nachrin2020social}. 

The Russian invasion of Ukraine on February 24, 2022, triggered a severe humanitarian crisis, resulting in a growing and urgent need for relief and recovery efforts for millions of people who remained in Ukraine, as well as for those who fled or are trying to leave the country as refugees. Social media networks have played a significant role in supporting the Ukrainian population, allowing citizens to share firsthand accounts of the conflict and fostering a sense of community and solidarity \cite{bang2023ukrainian}. These platforms have helped shape the narrative surrounding the conflict and galvanize support for those affected, further highlighting the importance of digital platforms during times of crisis \cite{balca2022who}.

\begin{figure}[t!]
\centering
\includegraphics[width=\columnwidth]{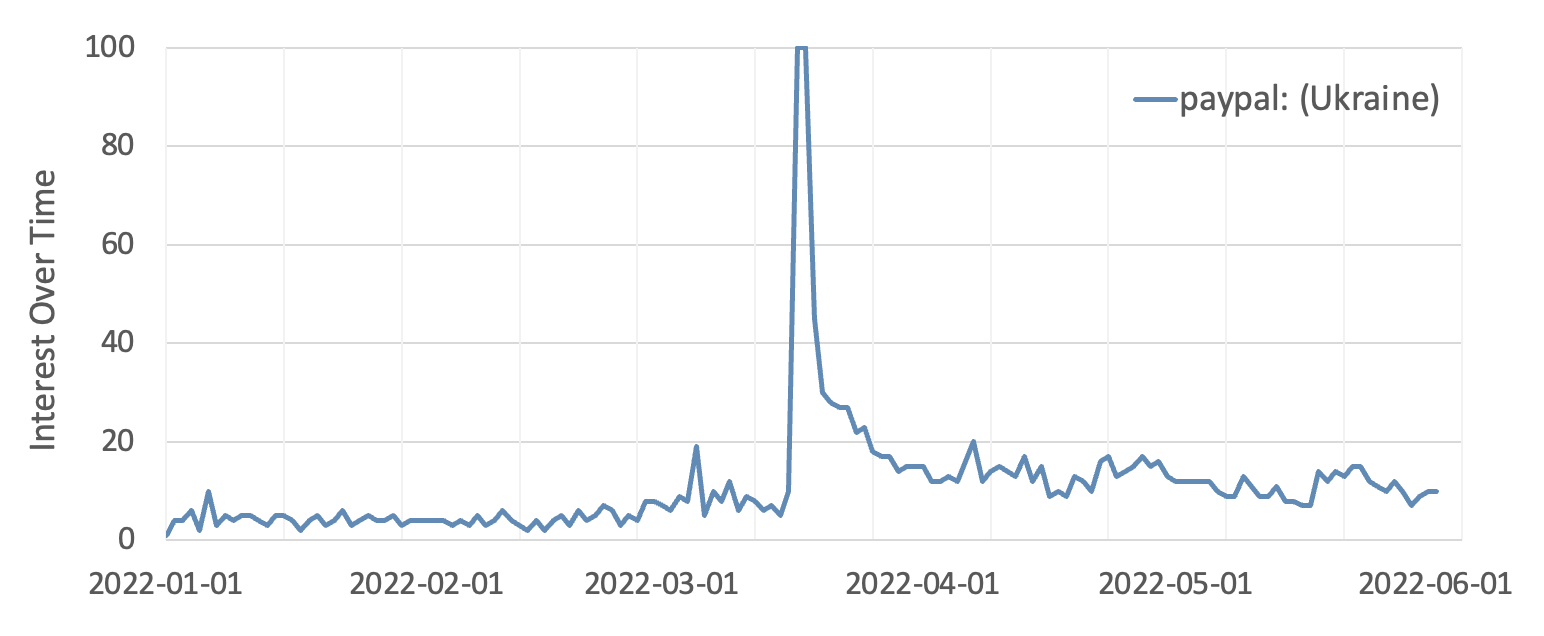} 
\caption{Volume of Google searches of the term \emph{Paypal} in 2022 (from Google Trends). An ``Interest Over Time" value of 100 represents the peak of popularity for the term under scrutiny. \vspace{-0.4cm}}
\label{paypal}
\end{figure}

Among the online sources of aid circulating on mainstream social media, crowdsourcing and crowdfunding platforms surfaced as prominent yet relatively novel means of support during the ongoing conflict. For example, a viral open-access \emph{Google Document} provides crucial, real-time information to 2.8 million people displaced by the war, including updates on border crossings, transportation, and support links.\footnote{https://www.forbes.com/sites/rashishrivastava/2022/03/15/how-one-google-doc-is-helping-thousands-of-ukrainian-refugees-navigate-borders/?sh=2f327a29d18e} \emph{PayPal} extended its service to facilitate peer-to-peer money exchange with no platform charges,\footnote{https://newsroom.paypal-corp.com/2022-03-Expanding-PayPals-Money-Services-To-Help-Ukraine-Humanitarian-Efforts} potentially resulting in a spike of Google searches (\textit{cf.} Figure \ref{paypal}). Furthermore, the crowdfunding and content creator platform \emph{Patreon} has helped independent Ukrainian journalists  and media outlets, such as \emph{Kyiv Independent}, raise millions of funds through grassroots crowdfunding.\footnote{https://www.ft.com/content/4a95520f-74ee-47f4-9d21-b3982e8512d8} 

Based on these premises, our study aims to investigate the online network of solidarity and mobilization during the Russian invasion of Ukraine.
We seek to understand the different types and diffusion patterns of online support resources that emerged during the conflict, while exploring the interplay between different social media platforms in disseminating support content. Using a mixed quantitative and qualitative approach, we also examine novel support practices on crowdsourcing and crowdfunding platforms. Specifically, we aim at answering the following research questions:

\begin{itemize}
    \item[\textbf{RQ1:}] \emph{How does the online network of support manifest and operate during the Russia-Ukraine conflict, and what is the relationship between social media platforms and online support sites?}
    \item[\textbf{RQ2:}] \emph{How prevalent are social media messages that lead to crowdsourcing and crowdfunding platforms, and how do they facilitate aid to the Ukrainian population?}
\end{itemize}

To answer these questions, we utilize a large-scale dataset comprising 68 million posts from four mainstream social media platforms: Twitter, Facebook, Instagram, and YouTube. Inspecting the URLs included in these posts (13.9M URLs across 520K web domains), we uncover a heterogeneous suite of support resources, including crowdsourcing and crowdfunding platforms as well as local and international funding organizations. This networked form of support leverages online channels to mobilize individuals and enable collective action, such as donating to online fundraisers, signing petitions, sharing information and resources, seeking and receiving assistance, and engaging in online activism.
Our research contributes to the existing body of knowledge by providing an overview of the interconnectivity between social media and support websites during crises, exploring the interplay between these platforms in the diffusion of support content, and identifying grassroots mobilization practices to aid the vulnerable population. 

\section{Related Work}

Current research that explore the role of social media networks during the Russia-Ukraine conflict can be grouped into two main categories. The first category reflects the negative aspects of online platforms, exploring how they are weaponized to spread misinformation, manipulate the public opinion, and sow division between different factions. The second category focuses on the positive facets of social media, and its beneficial role in sharing reliable and helpful information, mobilizing citizens, amplifying voices, and distributing humanitarian aid.


On the one hand, early investigations examined suspicious Twitter activity and identified groups of users promoting falsehoods and hate speech \cite{osomewp2, osomewp1, pierri2022does}. 
Leveraging a public dataset of over 63 million tweets, \citet{chen2022tweets} found evidence of public engagement with Russian state-sponsored media and other unreliable sources. 
A longitudinal study \cite{pierri2022propaganda} found that Russian propaganda on Twitter and Facebook decreased after the invasion, but misinformation persisted with verified accounts playing a key role in sharing low-credibility content. In this regard, a methodological framework based on pre-trained language models \cite{la2023retrieving} identified false and unsubstantiated claims related to the conflict circulating on Twitter in February 2022. 
\citet{geissler2022russian} suggested that pro-Russia Twitter messages primarily originated from India, South Africa, and the U.S., with bots playing a disproportionate role in their diffusion.
Finally, a multimodal dataset \cite{thapa2022multi} of text-image pairs of tweets demonstrated the prevalence of hate speech during political events and the usefulness of multimodal resources in detecting it. 

On the other hand, social media also played a positive impact in promoting social good and supporting populations affected by the conflict. 
An exploratory analysis \cite{carlsen2022ukrainian} presents preliminary findings on how social media was used for online mobilization of informal civic action to address the needs of Ukrainian refugees. 
Another study \cite{talabi2022use} found that social media storytelling helped Nigerian refugees in Ukraine receive help during the war, underscoring the importance of online platforms' communication infrastructure. 
An online ethnography study on Facebook \cite{carlsen2023some}, examined the entrance of a new group of activists into the refugee solidarity movement during the war and found that conflicts were rare and contained within different Facebook groups, despite ideological differences. 

Currently, relatively few studies have examined whether and how social media supports the vulnerable population affected by the Russian invasion of Ukraine, and most are qualitative in nature. Our study seeks to address this gap by leveraging large-scale social media datasets for quantitative analyses of online networks of solidarity and mobilization, thus adding to the literature on social media's positive role in promoting social good during crises and conflicts.

\section{Methodology}

\subsection{Data Collection}
The Russian invasion of Ukraine is a complex and rapidly evolving issue that has been extensively discussed across multiple media platforms. In this study, we collect social media posts related to the conflict from four platforms, namely Twitter, Facebook, Instagram, and YouTube, over a three-month period (February 22, 2022 - April 28, 2022), beginning two days prior to the invasion. The resulting dataset contains 68 million posts.

For what concerns Twitter data, we refer to an existing dataset \cite{chen2022tweets} collected through the \texttt{Standard v1.1 Streaming} endpoint\footnote{\url{https://developer.twitter.com/en/docs/twitter-api/v1}} by matching over 30 keywords (in English, Russian and Ukrainian language) related to Russia's invasion of Ukraine and identified with a snowball sampling approach; these keywords largely overlap with those specified in the Facebook and Instagram data collection. The resulting dataset contains 55.9M tweets.

We collect Facebook and Instagram data by leveraging CrowdTangle~\cite{crowdtangle}, a public tool that allows academics and journalists to search in the entire collection of posts shared by public pages and groups that have a minimum number of followers or that are added manually by other users of the platform.\footnote{For more details on the coverage see the official documentation: \url{https://help.crowdtangle.com/en/articles/1140930-what-data-is-crowdtangle-tracking}} We query the Crowdtangle API using a set of over 40 keywords (in English, Russian and Ukrainian language) related to the conflict and introduced in \cite{osomewp1,osomewp2}. The resulting dataset contains 11.8M posts. 

To collect data from YouTube, we utilize the YouTube API to extract video metadata from the YouTube videos shared on Twitter, Facebook, and Instagram. Our resulting dataset comprises metadata from over 275,000 videos.

For each post on Twitter, Facebook, Instagram, and YouTube, we identify, extract, expand, and parse any URL embedded in the shared messages or video descriptions. In total, posts from these mainstream platforms contain 13.9 million unique URLs across 520,000 web domains. Table \ref{dataset_summary} shows the number of posts, URLs, and web domains for each social media platform. 

\begin{table}
  \centering
  \small
  \begin{tabular}{cccc}
  \hline
    \textbf{Platform} & \textbf{Posts} & \textbf{URLs} & \textbf{Domains}\\
    \hline
    Twitter & 55,938,686 & 8,154,714 & 285,165\\
    Facebook & 11,777,025 & 5,542,833 & 185,567\\
    Instagram & 68,403 & 37,655 & 18,103\\
    Youtube & 275,937 & 204,344 & 30,405\\
    \hline
\end{tabular}
\caption{Number of posts, unique URLs, and web domains shared in every social media under analysis. \vspace{-0.2cm}}
\label{dataset_summary}
\end{table}


\subsection{Identifying Online Support Sources}

\begin{figure}[t]
\centering
\includegraphics[width=8cm]{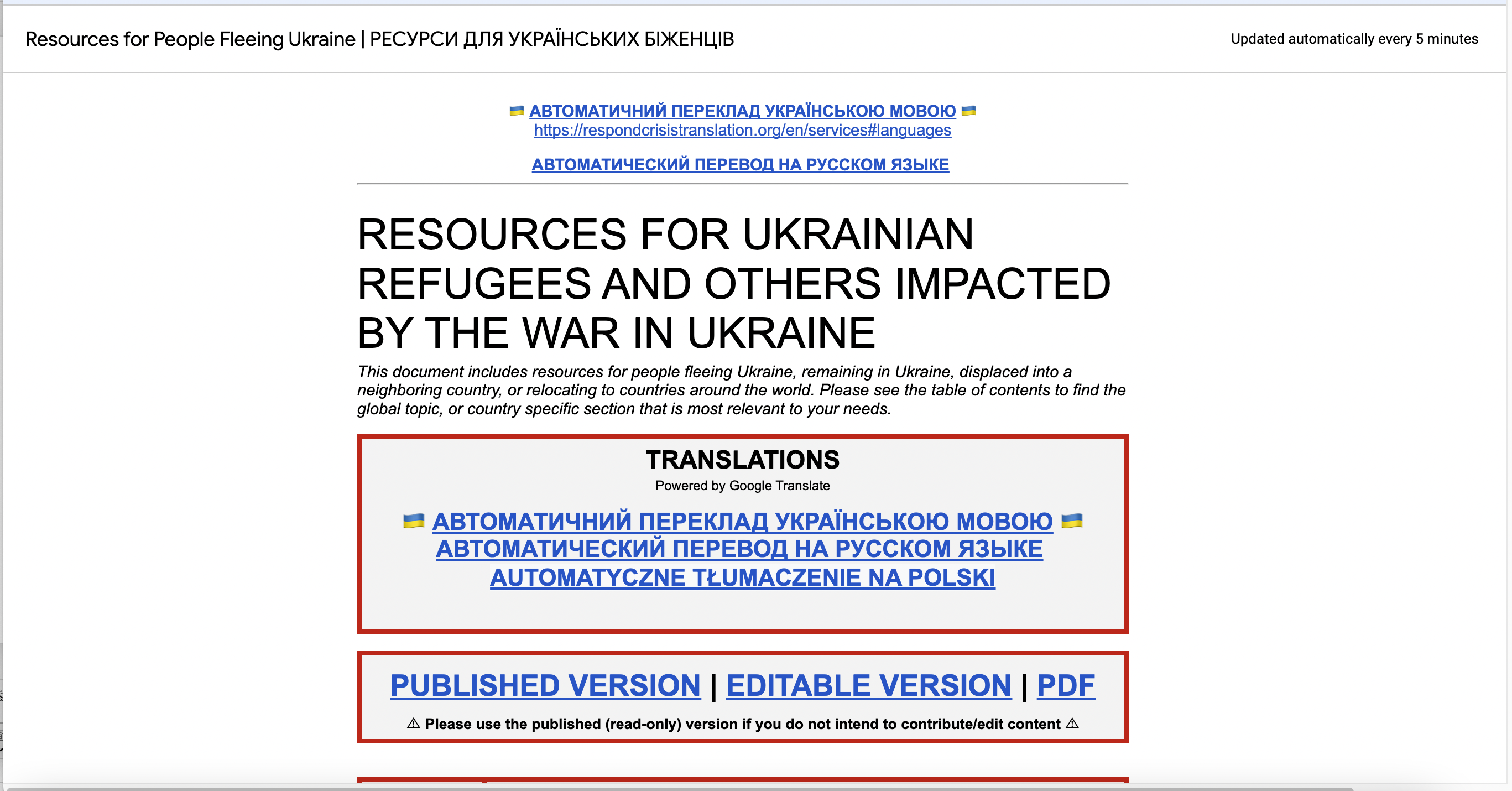} 
\caption{The Google Docs web-page titled ``Resources for People Fleeing Ukraine" is an example of crowdsourcing in action. The 89-page document offers real-time updates, including information on border crossings, housing, employment, medical support, and legal aid in neighboring countries. \vspace{-0.2cm}}
\label{example_googledoc}
\end{figure}

\begin{figure}[t]
\centering
\includegraphics[width=8cm]{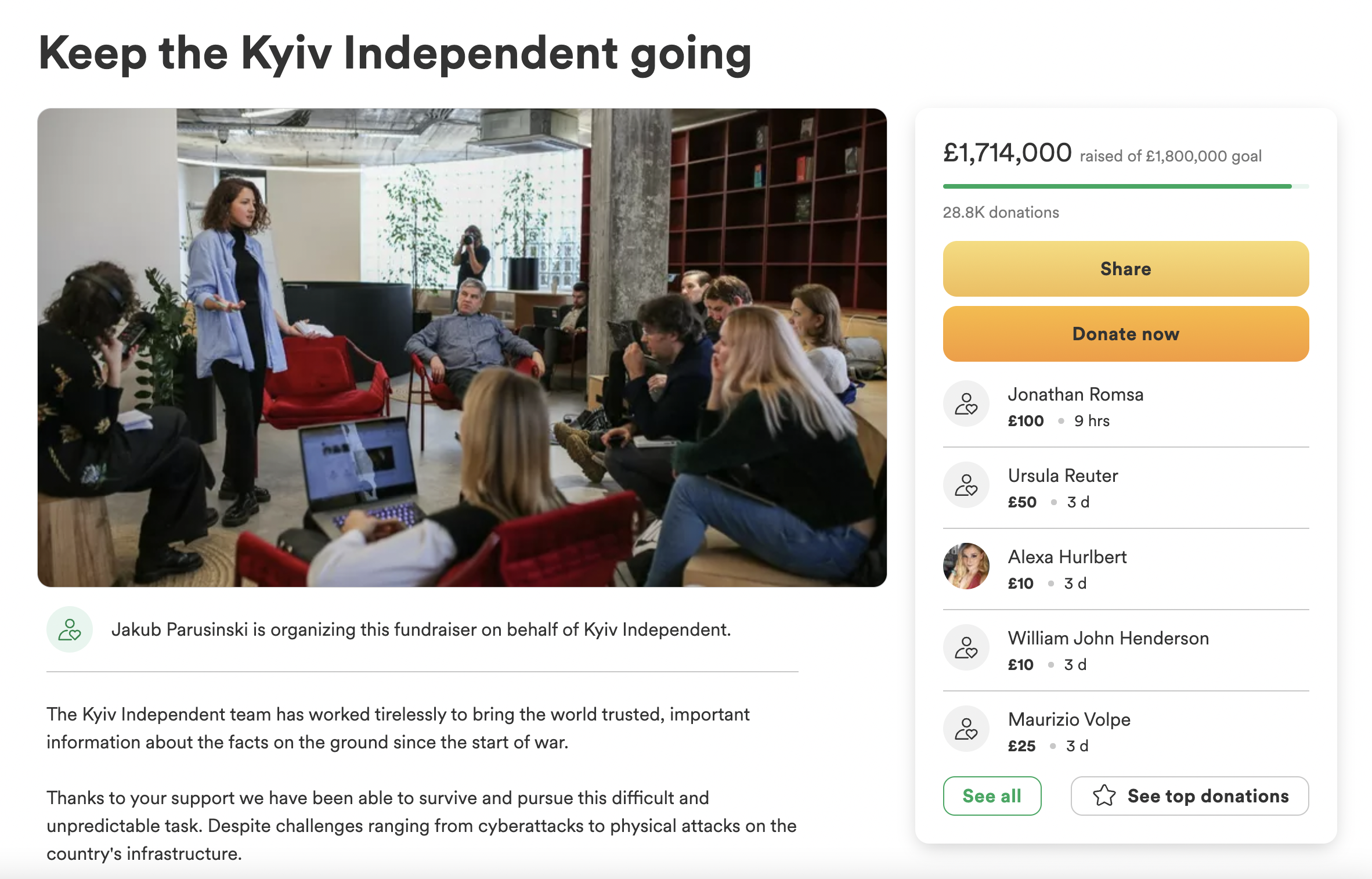} 
\caption{Example of the crowdfunding platform \emph{GoFundMe}, where the online journal \emph{Kyiv Independent} raised £1.7M in donations during the Russian invasion of Ukraine. \vspace{-0.4cm}}
\label{example_gofundme}
\end{figure}

This paper focuses on analyzing two forms of online support, namely crowdsourcing and crowdfunding, in response to the increasing use of related platforms as documented by reliable news outlets (\emph{cf.} the \emph{Introduction} Section). Crowdsourcing refers to soliciting contributions from a large group of people to obtain ideas, content, or services \cite{gao2011harnessing}, while crowdfunding involves raising funds for an entity by seeking small contributions from a broad population \cite{lu2014inferring}. Crowdsourcing and crowdfunding have proven to be effective in mobilizing support during crises such as natural disasters \cite{gao2011harnessing}, pandemics \cite{tully2019contextualizing}, and wars \cite{keatinge2019social} on social media platforms. Examples of crowdsourcing and crowdfunding webpages that were shared on social media platforms are presented in Figures \ref{example_googledoc} and \ref{example_gofundme}.

To create a comprehensive list of reliable sources of online support during the Russia-Ukraine conflict, we select donation venues recommended by highly credible media outlets (e.g., NPR, The Washington Post, CNBC), as well as information aggregation web-pages from trustworthy sources (e.g., the official website of War of Ukraine, PayPal’s “Support Ukraine Relief” page, the Ukrainian Institute London). To ensure comprehensive coverage across multiple social media platforms, we consider only the support sites that appeared on at least two of the four platforms analyzed. A complete list of the web domains used in this study is available in the \emph{Appendix} for interested readers.

Based on the observational analysis presented in the \emph{Results} section, we have found that Google Docs and Google Forms were extensively shared for crowdsourcing purposes. These tools are primarily used for collaborative document creation and information gathering. For instance, during the Russian invasion of Ukraine, a Google Doc\footnote{bit.ly/40wqdc0} was created to document labs around the world that were willing to accept Ukrainian scientists. Similarly, a Google Form\footnote{https://forms.gle/2YG53qWEUEZWaxH48} was used to collect links to digitized collections of Ukrainian cultural heritage to protect them against destruction. The collaborative nature of these tools enables the swift mobilization of resources and support from a large number of individuals, making them an effective means for crowdsourcing during times of crisis \cite{lai2017study}.


For what pertains to fundraising sites, we further specified four categories based on the funding mechanics: 
\begin{itemize}
    \item Crowdfunding platforms: Digital platforms that allow individuals or organizations to create a fundraising campaign, usually charging a fee based on the funds raised, e.g., Patreon, PayPal, GoFundMe, Kickstarter.
    \item Ukraine government fundraising sites, e.g., National Bank of Ukraine (bank.gov.ua), Ministry of Health of Ukraine (moz.gov.ua).
    \item Local NGOs/funds: Nongovernmental organizations and charity funds that operate in Ukraine local level, e.g., Come Back Alive (savelife.in.ua), Razom Emergency Response (razomforukraine.org), Serhiy Prytula Charity Foundation (prytulafoundation.org).
    \item International NGOs/funds: Nongovernmental organizations and charity funds that operate internationally, e.g., UN Refugee Agency (unhcr.org), CARE (care.org), GlobalGiving Foundation (globalgiving.org), International Committee of the Red Cross (icrc.org).
\end{itemize}

This categorization is informed by previous work \cite{carlsen2022ukrainian}, which emphasized the important role that both formal and informal civil societies play during the Ukraine refugee crisis: When state institutions and local/regional NGOs were unable to provide immediate assistance, informal voluntary support networks for Ukraine organized on social media emerged to fill the gap \cite{carlsen2022ukrainian}. The final list yields 51 support site domains in total.

\section{Results}
\subsection{RQ1: Examining the Network of Support}

To identify and characterize the online network of support during the Russia-Ukraine conflict, we examine the interconnectivity between social media platforms and support sites. We aim at providing a better understanding of the spread of information across different platforms and how this affects the online support landscape.

We initiate our analysis by inspecting posts from Twitter, Facebook, Instagram, and YouTube that link to the 51 support sites described earlier. 
We note that only support sites embedded in social media posts are considered, excluding thus indirect mentions circulated through private messages, images, and videos. Therefore, the assessed prevalence of online support should be regarded as a lower-bound estimate in this case. 
Table \ref{supoort_site_count} displays the number of support site URLs and domains shared on each platform. 

Our analysis indicates that Twitter has the highest number of shared support URLs, making up 77.10\% of all support URLs across the four social media platforms, but Instagram and YouTube have a relatively higher percentage of support URLs shared, accounting for 2.51\% and 1.62\% within their respective platforms.
It is worth noting that the vast majority (over 90\%) of the 51 support sites identified in our analysis were shared across all four social media platforms, indicating their potential to serve as a comprehensive source of help, support, and aid.

\begin{table}[t!]
  \centering
  \small
  \begin{tabular}{ccc}
  \hline
    \textbf{Platform} & \textbf{no. of support URLs} & \textbf{no. of support sites}\\
    \hline
    Twitter & 295,467 (0.53\%) & 50\\
    Facebook & 60,151 (0.51\%)& 51\\
    Instagram & 1,717 (2.51\%)& 46\\
    Youtube & 25,920 (1.62\%)& 49\\
    \hline
\end{tabular}
\caption{Number of support URLs and domains shared in each social media. The percentage represents the proportion of support URLs over the total number of URLs shared in the platform. \vspace{-0.2cm}}
\label{supoort_site_count}
\end{table}

\begin{figure}[t!]
\centering
\includegraphics[width=\columnwidth]{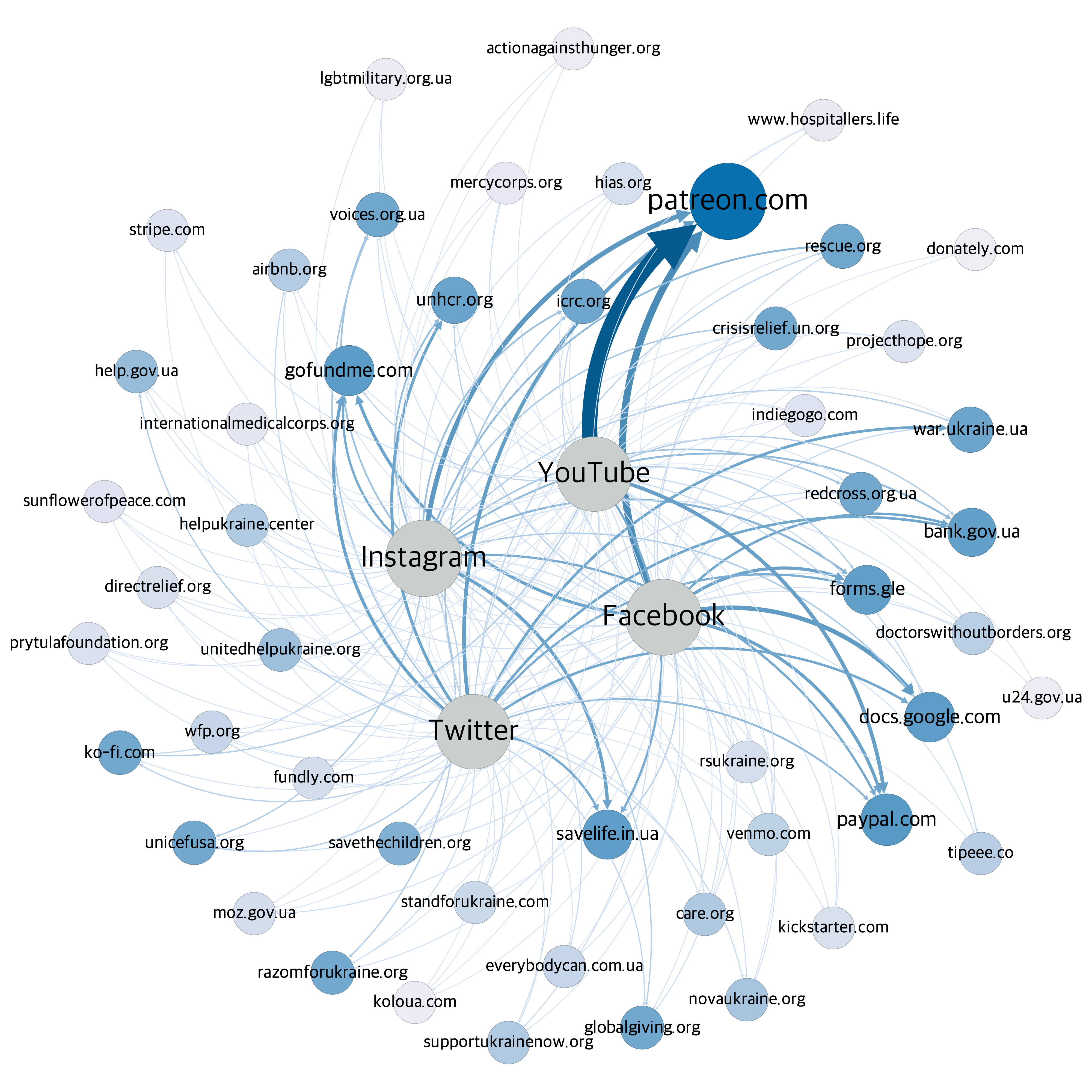} 
\caption{The online network of support: Connections between Social media platforms and support sites are represented with normalized, directed, and weighted edges. \vspace{-0.4cm}}
\label{network-viz}
\end{figure}

To examine the online network of support during the Russia-Ukraine conflict, in Figure \ref{network-viz}, we portray the connections between social media platforms and support sites.
The network graph is presented in a circular layout, where nodes representing social media platforms are placed centrally in grey with fixed sizes.
The size and color of the nodes representing support sites are based on their weighted in-degree, which is determined by the number of links pointing to them. Thus, larger and darker nodes indicate support sites that have more incoming links. 
The edges in the graph are directed and weighted, pointing from the \emph{source} social media platform to the \emph{target} support site.
The width and color of the edges represent the number of connections between the source and the target, with darker and wider edges indicating more frequent links. It is important to note that the edge's weight is normalized for each platform based on the total number of shared links to support sites.


Taking all four social media platforms into account, the top 10 most shared support sites are as follows (see the \emph{Appendix} for more details): Patreon, GoFundMe, Google Docs, UN Refugee Agency, National Bank of Ukraine, the Ukrainian Official Website of the war (war.ukraine.ua), Google Forms, PayPal, Come Back Alive, and the International Committee of the Red Cross. 

Based on Figure \ref{network-viz}, two interesting patterns can be observed. First, the majority of URLs shared on YouTube (62.30\%) are linked to Patreon, which is a membership-based platform that provides financial support to content creators, including journalists, and is known for supporting independent journalism through direct and recurring financial support \cite{hunter2016s}.


Second, the distribution of links from Twitter and Facebook appears to be more evenly spread across various crowdfunding and crowdsourcing sites, whereas links from Instagram and YouTube tend to point towards specific crowdfunding platforms such as Patreon and PayPal. 
It is probable that Instagram and YouTube have limited potential to disseminate crowdsourcing sites and donation links, as they are mainly visual-based social media platforms without integrated features to upload or embed URLs and documents. Nonetheless, users can still share links by including them in their post captions.
Contrarily, Twitter and Facebook allow users to embed Google Docs and Forms directly into their posts, making it more accessible to a wider audience and increasing engagement with the document.



\begin{figure}[t]
\centering
\includegraphics[width=7.5cm]{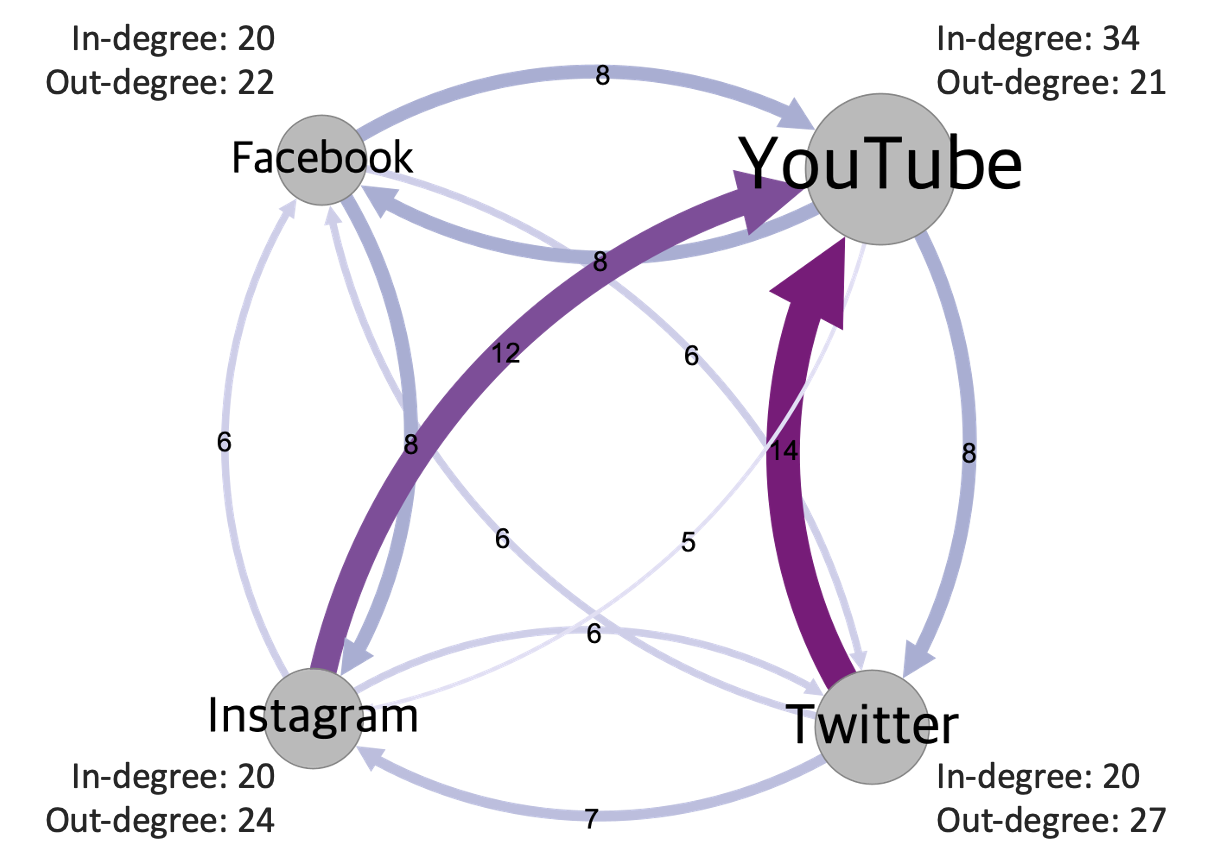} 
\caption{Network visualization of pairwise Granger causality tests between social media networks. \vspace{-0.4cm}}
\label{granger_network}
\end{figure}

To gain a better understanding of the interplay between social media platforms and support sites, we conduct pairwise Granger causality tests on the volume of posts linking to different support sites shared on the four social media networks. Specifically, we extract the number of links to every support site diffused daily on the four social media platforms within the observation period. 
We then perform the Granger causality test for each support site separately, considering every pair of social media networks. Before the test, we detrend the corresponding time series, setting a maximum lag parameter of 3 days and using a significance level threshold of \emph{p-}value $<$ .05, following previous studies on social media temporal analysis \cite{kuzma2022influencing, park2017information}. It is important to note that we only examine instances when a support site was linked at least 10 times, in order to exclude cases where support sites were only rarely linked.


In Figure \ref{granger_network}, we present a network diagram generated using the statistically significant results of the pairwise Granger causality test.
The diagram shows the relationships between different social media networks. The size of each node is proportional to the weighted degree, and the weight of the edges indicates the number of support sites where the Granger test showed statistically significant Granger-causality.
For instance, a directed edge with a weight equal to $n$ from Facebook to YouTube indicates that Facebook Granger-causes YouTube in the diffusion of $n$ support sites (the interested reader can refer to the Appendix for the detailed results of every Granger causality test).

\begin{figure}[t]
\centering
\includegraphics[width=\columnwidth]{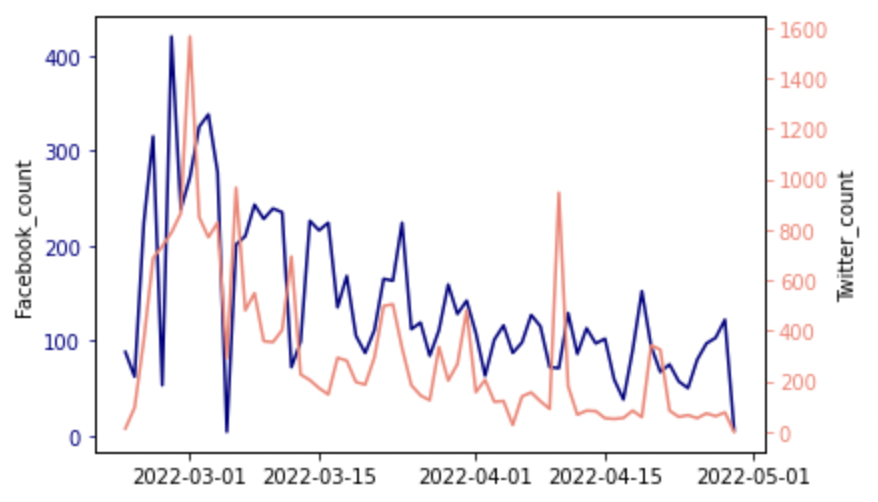} 
\caption{Daily volume of posts linking to Google Docs on Facebook (blue) and Twitter (pink). In this example, Facebook Granger-causes Twitter with a lag of 3 days and a \emph{p-}value $<$ .05. \vspace{-0.2cm}}
\label{granger_case}
\end{figure}
From Figure \ref{granger_network}, two facts are worth noting. First, Twitter has the largest out-degree, which suggests it Granger-causes the diffusion of support sites more than the other three platforms. Specifically, Twitter Granger-causes both YouTube and Instagram in sharing links to crowdsourcing sites, representing the main source of this kind of support. However, Facebook Granger-causes Twitter when considering the diffusion of Google Docs, as shown in Figure \ref{granger_case}. Second, YouTube appears to have the largest in-degree, suggesting that support on YouTube is mainly triggered by other platforms. Interestingly, most of these instances regard crowdfunding platforms, even though YouTube Granger-causes both Twitter and Facebook in sharing the crowdfunding site PayPal. Finally, our results show that Facebook mainly Granger-causes Instagram in crowdfunding sites.

\subsection{RQ2: Characterizing Online Support Sites}

Building upon the findings of RQ1, we delve deeper into the dissemination of crowdsourcing and crowdfunding URLs, examining their scale, scope, and content. Specifically, to address RQ2, we select Google Docs and Google Forms as examples of crowdsourcing websites, and the list of identified crowdfunding platforms for the crowdfunding sites. The rationale behind this selection is that our study places greater emphasis on grassroots mobilization efforts led by individuals and communities during social crises, rather than those implemented by government, local, and international organizations. Table \ref{two_support_forms} provides an overview of the number of posts containing URLs directing to crowdsourcing and crowdfunding platforms on each social media platform.


\begin{table}[t!]
  \centering
  \small
  \begin{tabular}{ccc}
  \hline
    \textbf{Social Media} & \textbf{Crowdsourcing} & \textbf{Crowdfunding}\\
    \hline
    Twitter & 38,362 (12.98\%) & 89,260 (30.21\%)\\
    Facebook & 15,483 (25.74\%) & 29,314 (48.73\%)\\
    Instagram & 137 (7.98\%) & 554 (32.27\%)\\
    Youtube & 910 (3.51\%) & 21,196 (81.77\%)\\
    \hline
\end{tabular}
\caption{Number of URLs to crowd-sourcing/-funding platforms shared on each social media. The percentage shows the proportion of crowd-sourcing/-funding URLs over all the  shared support URLs (percentages do not sum up to 100\% as government, local, and international fundraising URLs are not considered).  \vspace{-0.4cm}}
\label{two_support_forms}
\end{table}



To examine the content shared within crowdsourcing and crowdfunding URLs, we use a mixed qualitative and quantitative approach.
We select two samples: the top 100 most shared Google Docs and Forms on each platform (accounting for 57.61\% of the total number of shares of Google Docs and Forms), and the top 100 most shared crowdfunding platform URLs on each platform (accounting for 29.30\% of the total number of shares of crowdfunding platform URLs). 
The list of all analyzed URLs is released to the research community to enable further research and support to the Ukrainian population.\footnote{https://github.com/angelayejinyi/russia-ukraine-network-support}


\paragraph{Crowdsourcing Sites}
Crowdsourcing actions provide an organized approach to reporting and addressing aid requests in times of crisis. \cite{conrad2020improving}.
The features provided by Google Docs and Forms, such as real-time synchronization, version control, and easy document sharing, can be advantageous for facilitating coordination and information sharing in humanitarian relief efforts \cite{liu2014crisis}. In the top 100 shared Google Docs and Forms URLs across four platforms, we manually coded nine distinct types of support: 

\begin{itemize}
    \item General donation: All forms of donations except for cryptocurrency donations, e.g., ``Donate To Locals In Ukraine", ``Stand with Ukraine Bundle".
    \item Non-Fungible Tokens (NFT) donation: Selling NFTs as a form of crypto donation, e.g., ``Refugees Token | Airdrop - Donation", ``CREATE FOR UKRAINE".
    \item Petition, appeal, open letter: Signing open letters to condemn the military action and call for support, e.g., ``Open letter to the president of the russian federation", ``Sign to support the International Appeal to NATO".
    \item Resources \&  information: Helpful resources and information aggregation pages, e.g., ``DAT | Ukraine Resources \& Information", ``Free Cyber \& Humanitarian Services for Ukraine"
    \item Professional support: Support for professionals of various fields in Ukraine, e.g., gaming industry professionals, lab scientists, artists, and cultural workers.
    \item Foreigner support: Support for foreigners in Ukraine.
    \item Pet and animal support: Support for pets and animals in Ukraine, e.g., ``Dog Adoption Knowledge Questionnaire".
    \item Refugee sign-up: Ukrainian refugee signup forms, e.g., ``I need accommodation in Madrid".
    \item Activity \& volunteer registration: Activities, events, and volunteer recruitment to support Ukraine, e.g., ``Stop promising, start acting! Street action to support Ukraine".
\end{itemize}

\begin{figure}[t]
\centering
\includegraphics[width=\columnwidth]{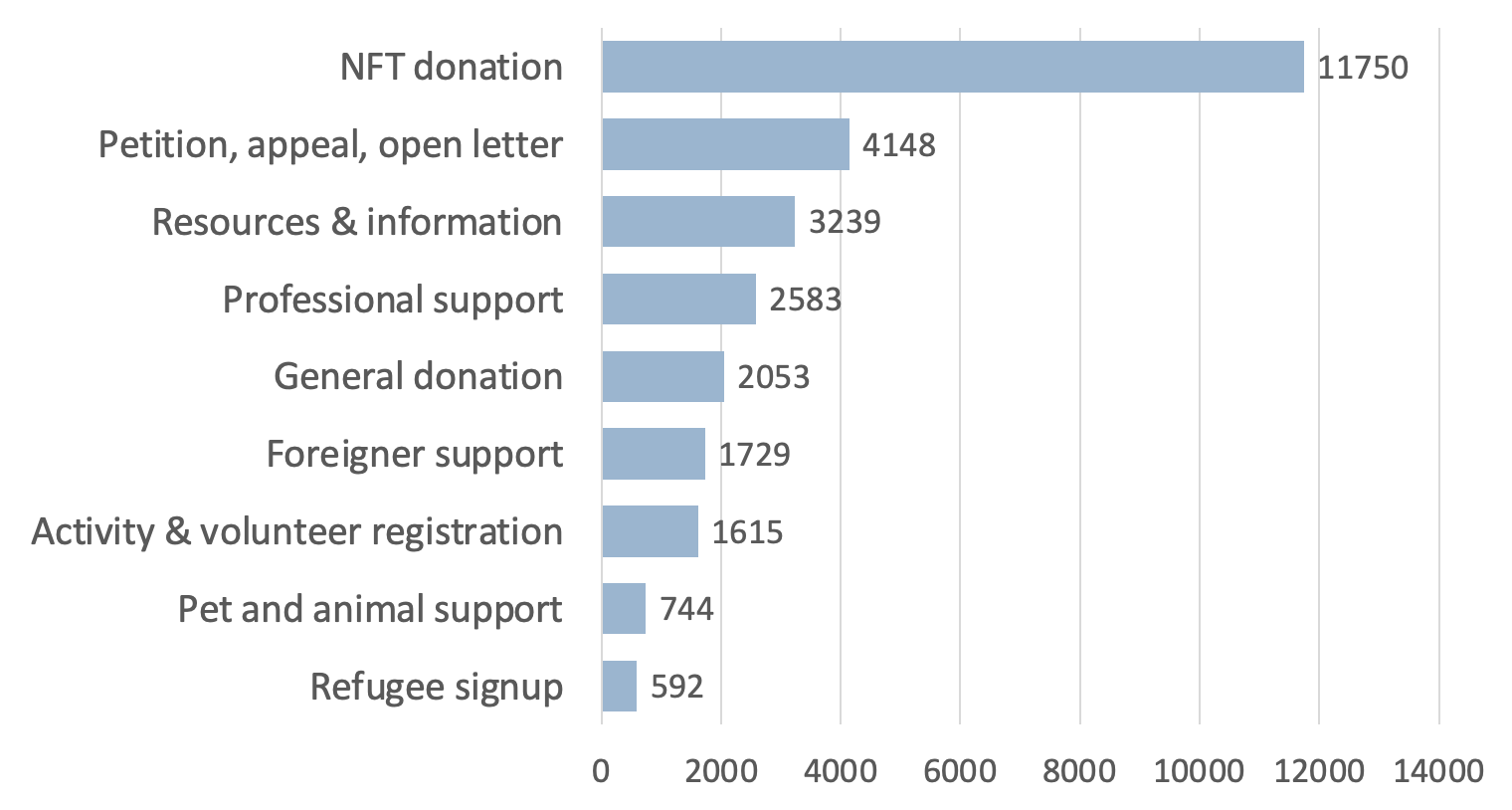} 
\caption{Number of posts sharing Google Docs and Forms related to each category on the four social media networks. \vspace{-0.2cm}}
\label{topic_summary}
\end{figure}

\begin{figure}[t]
\centering
\includegraphics[width=\columnwidth]{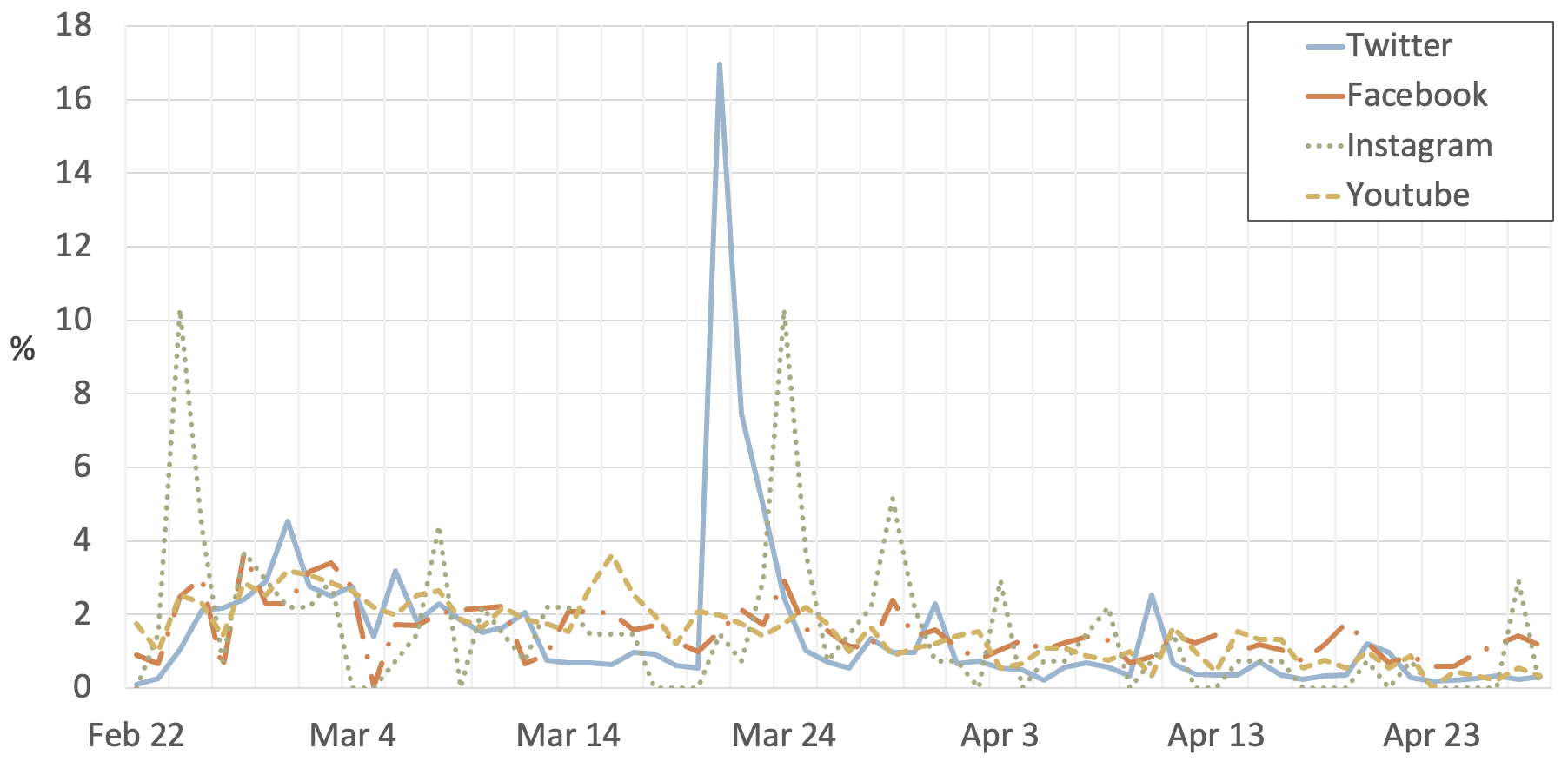} 
\caption{Daily proportion of posts sharing URLs to Google Docs and Forms on each social media network. 
\vspace{-0.4cm}}
\label{google_docs_daily}
\end{figure}

The number of posts sharing each topic is shown in Figure \ref{topic_summary}, whereas the daily volume of links to Google Docs and Forms on each social media platform are displayed in Figure \ref{google_docs_daily}. It can be concluded that ``NFT donation", ``Petition, appeal, open letter", and ``Resources \& information" are the most common forms of support via crowdsourcing.
Interestingly, a significant amount of Google Docs and Forms are related to NFT and cryptocurrency donations, which has emerged as a popular crowdfunding approach during the Russia-Ukraine conflict. NFTs are unique digital assets that are verified on a blockchain, providing proof of ownership and scarcity. They allow for a more decentralized and democratic approach to crowdfunding, enabling anyone to participate in supporting a project without intermediaries or gatekeepers \cite{chohan2017decentralized}. Although NFT donations appear to be a new form of solidarity, we should consider that this can potentially fall into less benevolent activities, including spamming, scams, and cryptocurrency manipulations \cite{nizzoli2020charting, pierri2022does}.
Also, the peak related to the Twitter trend in Figure \ref{google_docs_daily}, on March 21, 2022, is mainly related (6,131 URLs, 94.29\% of the URLs shared that day) to a Google Form\footnote{bit.ly/40E7NWS} named ``Refugees Token | Airdrop - Donation", offered by a charity project called Refugees Token (RFG), which allows refugees to send funds to their families through cryptocurrencies.




Besides the surge in NFT donations, we observe that the usage of Google Docs and Forms is diverse and multifaceted. Petitions and appeals are often shared through these sites, indicating their potential as a platform for advocacy and activism. Similarly, the dissemination of important resources and information such as shelters, transportation, food distribution locations, resource distribution, and medical aid is another was strategically carried out through these collaborative tools. Finally, Google Docs and Forms were also utilized to coordinate support efforts for different groups affected by the crisis, including professionals, foreigners, and animals. This demonstrates the versatility of Google Docs and Forms in facilitating various types of support efforts during the Russia-Ukraine conflict.

\begin{figure}[t]
\centering
\includegraphics[width=\columnwidth]{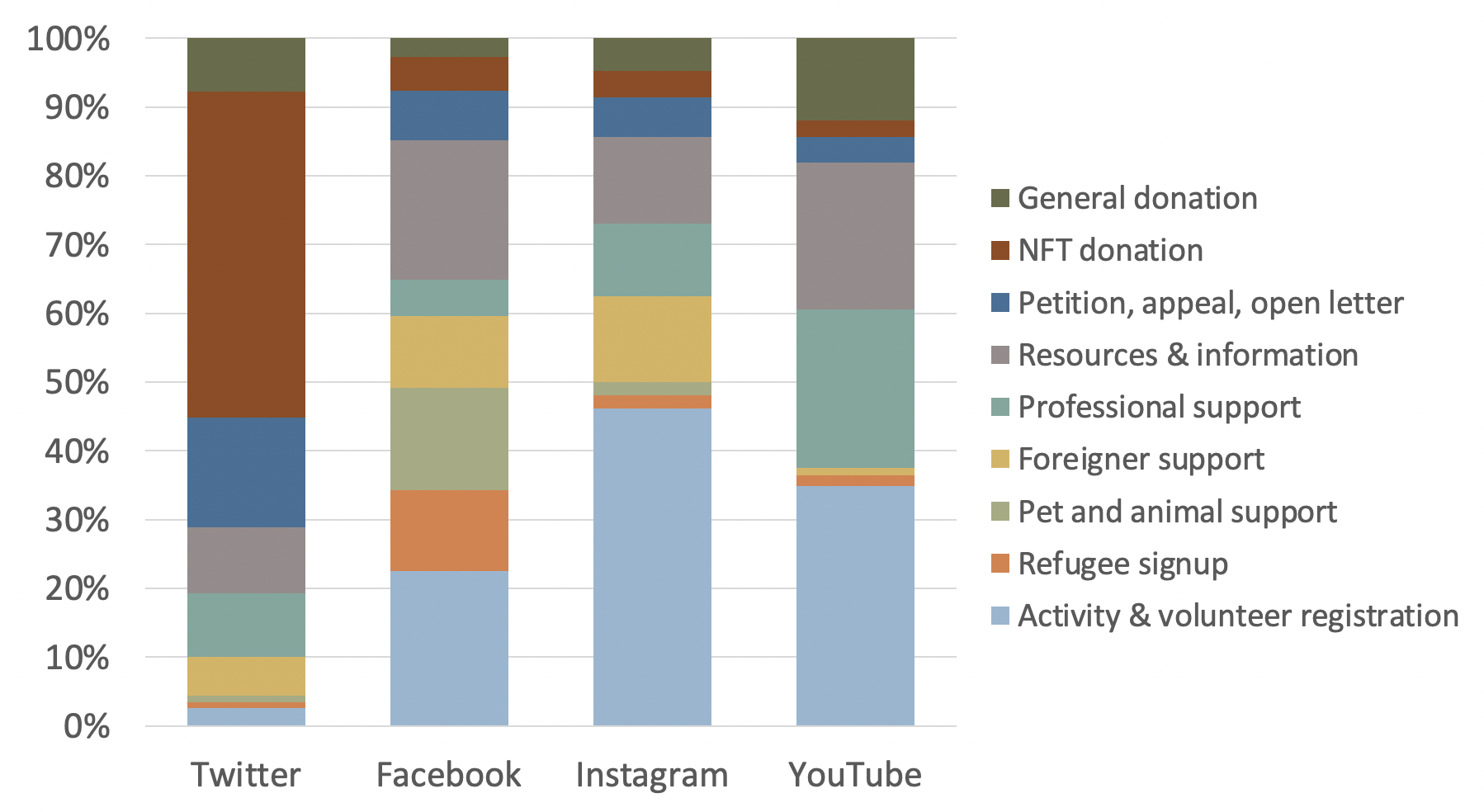} 
\caption{Category distribution of Google Docs and Forms on each social media network. \vspace{-0.3cm}}
\label{google_docs_distribution}
\end{figure}

Figure \ref{google_docs_distribution} displays the distribution of the nine categories of support on each platform. We observe that Twitter and Facebook had a more heterogeneous set of forms of support if compared to Instagram and YouTube, which had a particular focus on activity and volunteer registration. Interestingly, NFT donations are particularly relevant on Twitter, while Facebook users are active in signup activities and animal support.

\begin{figure}[t]
\centering
\includegraphics[width=\columnwidth]{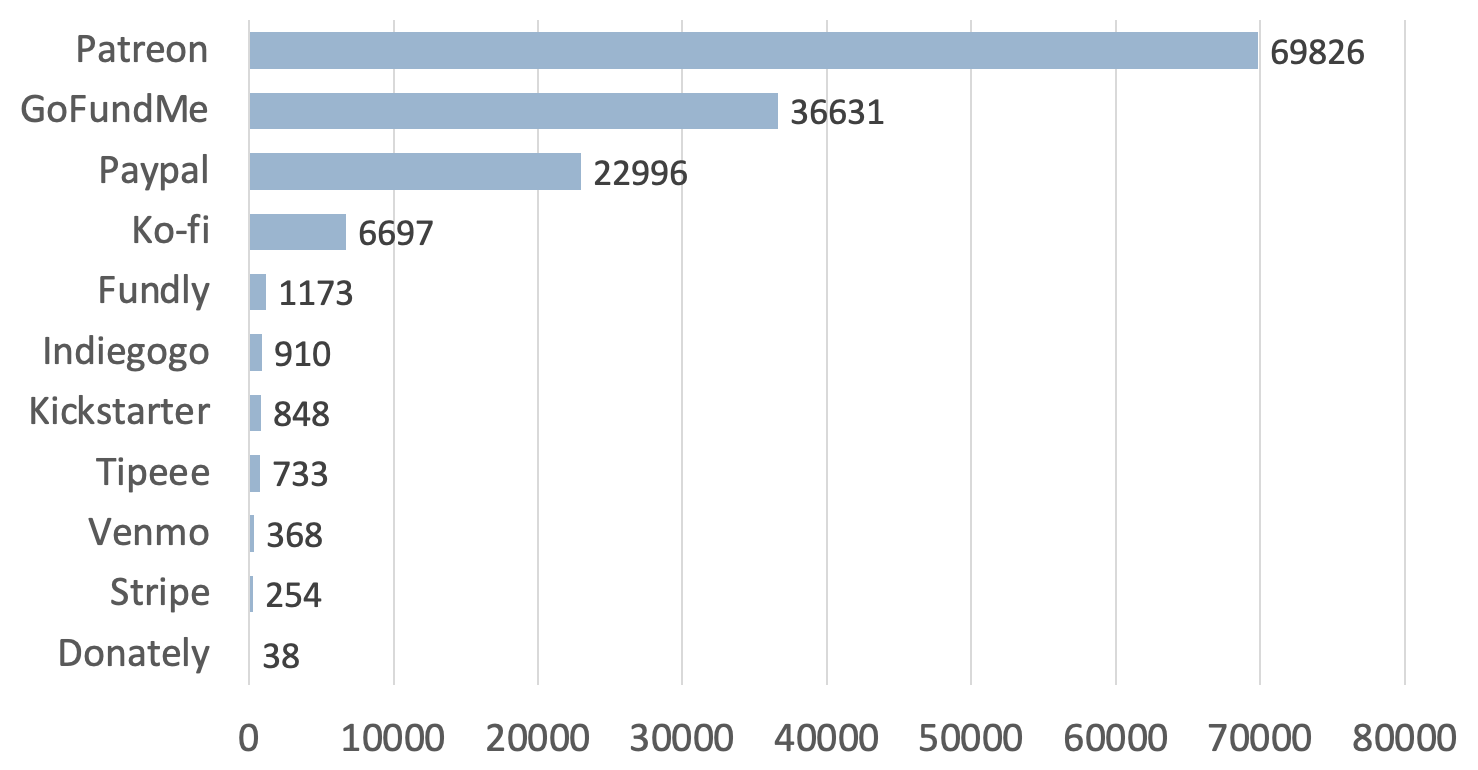} 
\caption{Distribution of crowdfunding platform URLs on each social media network. \vspace{-0.4cm}}
\label{crowdfunding_summary}
\end{figure}

\begin{figure}[t]
\centering
\includegraphics[width=\columnwidth]{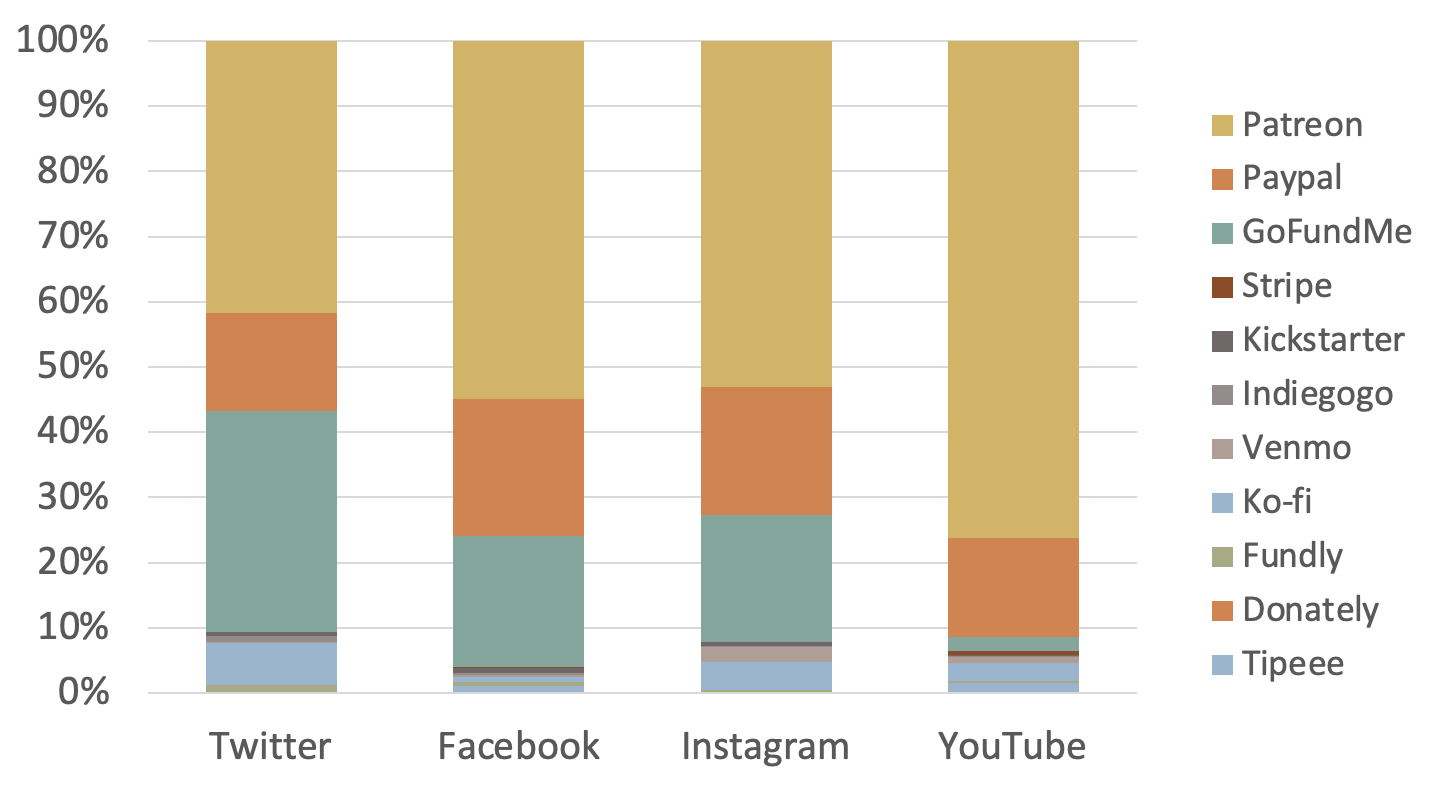} 
\caption{Number of social media posts sharing URLs to major crowdfunding platforms. \vspace{-0.3cm}}
\label{crowdfunding_distribution}
\end{figure}

\paragraph{Crowdfunding Platforms}
During crises, people without sufficient resources often rely on their social connections for aid, as official support may be delayed or inaccessible \cite{ho2021influence}. 
According to \citet{ho2021influence}, crowdfunding provides an alternative by allowing anonymous users to donate directly to fundraisers, giving underserved populations access to wider social networks for financial support. 
Crowdfunding platforms like Patreon, GoFundMe, and PayPal have gained popularity as a means to provide online financial assistance during the Russia-Ukraine conflict. Figures \ref{crowdfunding_summary} and \ref{crowdfunding_distribution} show the number of URLs shared across the four social media networks for each crowdfunding platform. It can be concluded that Patreon, GoFundMe, and PayPal are the most commonly used crowdfunding platforms across all the analyzed social media networks.


\begin{figure}[t]
\centering
\includegraphics[width=\columnwidth]{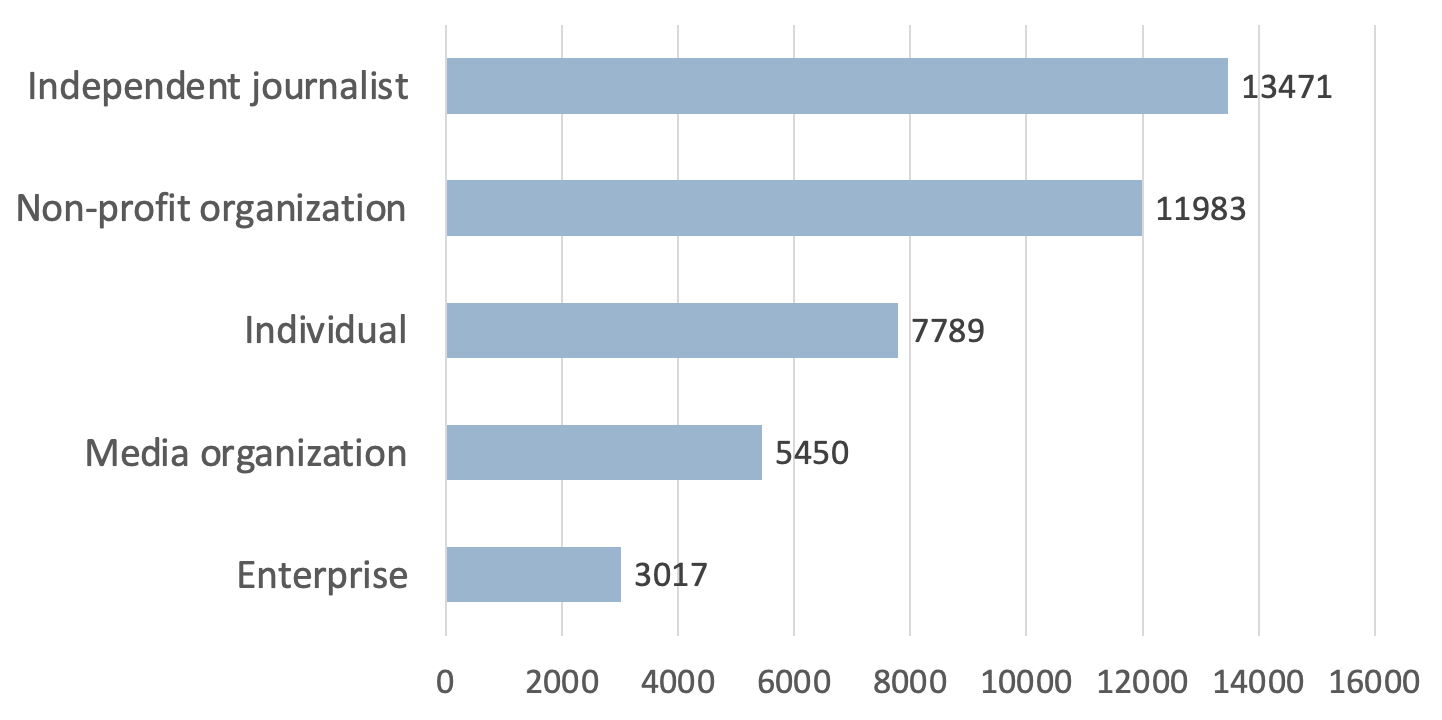} 
\caption{Number of posts to crowdfunding URLs based on the type of fundraising entity.\vspace{-0.4cm}}
\label{fundraising_entity}
\end{figure}

\begin{figure}[t]
\centering
\includegraphics[width=\columnwidth]{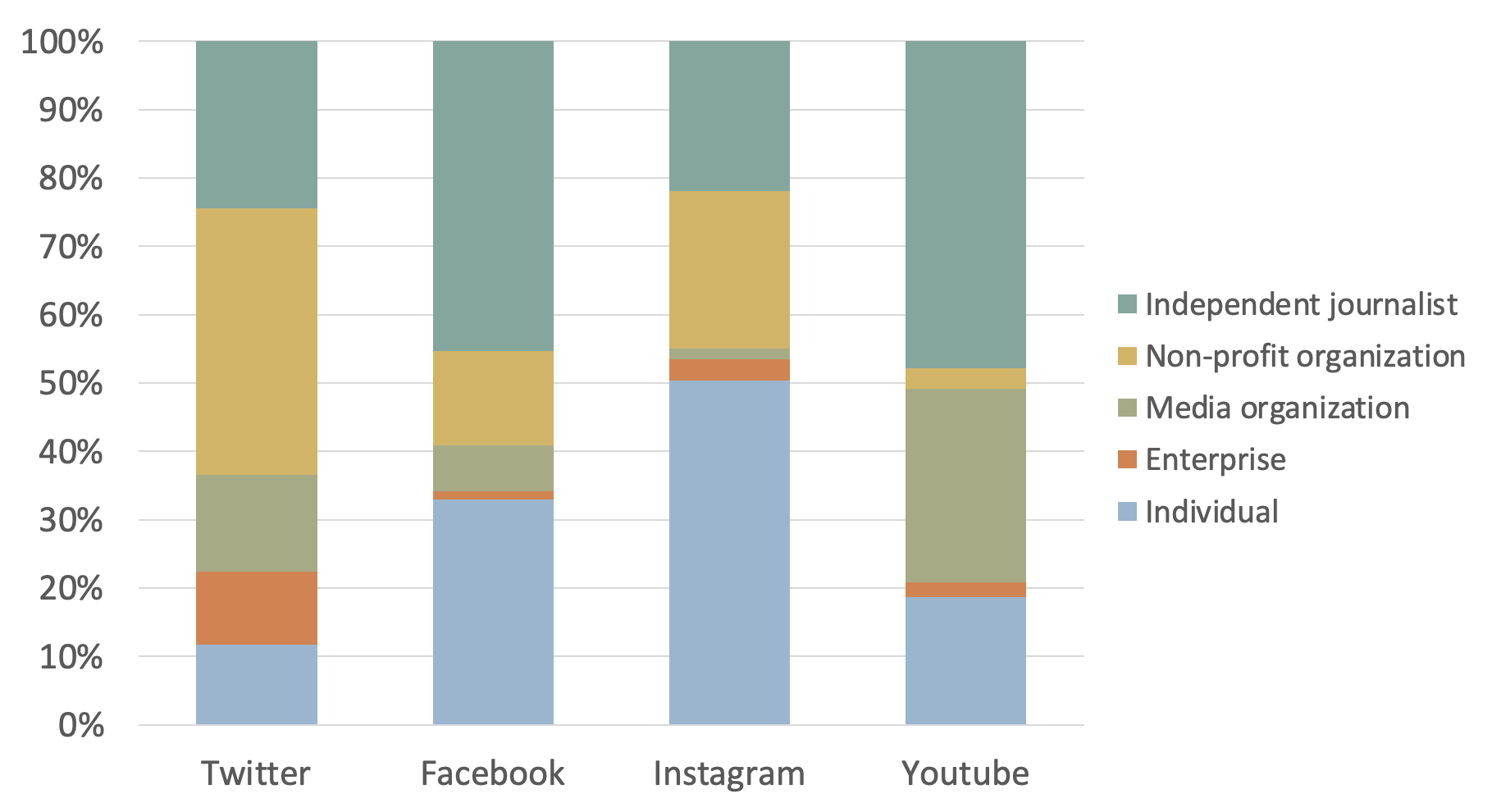} 
\caption{Distribution of fundraising entities on each social media network. \vspace{-0.4cm}}
\label{entity_distribution}
\end{figure}

Further, we characterize the various types of funded entities on these crowdfunding platforms. We identify five distinct entities in the top 100 most shared crowdfunding URLs across four social networks, including independent journalists, media organizations, non-profit organizations, and individual and enterprise fundraisers. Figure \ref{fundraising_entity} shows the overall distribution of posts linking to different crowdfunding entities, while Figure \ref{entity_distribution} shows their distribution within each social platform. 
One noteworthy observation is the significant presence of independent journalists and media organizations utilizing crowdfunding platforms for financial support. As they lack financial resources and backing from larger organizations, these crowdfunding platforms provide a way for creators to connect directly with their audience and receive financial support for their work. Notable examples include Kyiv Independent, Bellingcat, Malin Bot, John Sweeney, and Alexei Navalny’s team. Interestingly, our analysis reveals that 76.91\% (14,552 URLs) of crowdfunding links initiated by independent journalists and media organizations originate from Patreon. 
Additionally, the proportion of links from YouTube to Patreon is the highest among the four social media networks analyzed (76.18\%). This may be due to the platform's capability to provide support to content creators through monthly subscriptions.

Also, we note that many non-profit organizations are utilizing crowdfunding platforms, such as GlobalGiving, Come Back Alive, The Fix Foundation, and United Help Ukraine, for fundraising purposes during the Russia-Ukraine conflict, particularly on Twitter and Instagram. These platforms offer an additional and effective way for non-profits to tap into a broad base of potential supporters. As some medium or small-sized organizations may not have a significant media presence or large following, crowdfunding platforms often provide donors with detailed information about the non-profit and its mission, which can help build trust and encourage donations. Finally, individual fundraisers account for a significant proportion of crowdfunding URLs on Facebook (31.84\%) and Instagram (46.36\%). This is mainly due to the presence of public figures in Ukraine, such as politician Serhiy Prytula and author Oleksandr Paliĭ, who make up the majority of URLs. It is also common to find ordinary people acting as grassroot fundraisers on platforms like GoFundMe (38.66\%), PayPal (27.91\%), and Patreon (21.77\%).

\section{Conclusions}
\subsection{Findings and Contributions}
This study aims to analyze the interconnectivity between social media platforms and support sites during the Russia-Ukraine conflict. 
Using network-based visualization and Granger-causality analysis, we found that the observed social media networks---Twitter, Facebook, Instagram, and YouTube---differently contribute to the spread of support sites. Indeed, the links shared on Twitter and Facebook are more evenly distributed across various crowdsourcing and fundraising websites, while Instagram and YouTube tend to focus more on crowdfunding platforms such as Patreon and PayPal. 
Leveraging a combination of qualitative and quantitative methods, we investigate social media posts containing links to crowdsourcing sites and crowdfunding platforms during the Russia-Ukraine conflict. 
On the one hand, our findings show that Google Docs and Forms are commonly used for crowdsourcing various types of support, particularly for NFT and cryptocurrency donations, petitions, and online resources and information. 
On the other hand, Patreon, GoFundMe, and PayPal are identified as the main crowdfunding platforms, which provide financial assistance to people in need, as well as independent journalists, alternative media, and non-profit organizations.

Our study offers several contributions to the understanding of the role of social media platforms and support sites during a crisis. 
First, it sheds light on the ways in which social media platforms and support sites are interconnected during a crisis, revealing the interplay of different social media platforms in support diffusion. This adds to the growing body of quantitative research on social media as a force for social good in distressed environments. 
Second, we highlight the potential of crowdsourcing and crowdfunding in enabling grassroots mobilization.
We found that
crowdsourcing facilitates large-scale content production and consumption, creating a comprehensive, collective resource for disaster relief efforts. 
Crowdfunding, instead, supports vulnerable communities in a timely and efficient manner, as well as democratizes news production by providing a direct source of income for independent journalists and alternative media.
Our research calls for further investigations into these benevolent activities to promote positive impact and prosocial behavior through social media.

\subsection{Limitation}
Our study has several limitations that should be acknowledged. First, the dataset is highly imbalanced, with Twitter and Facebook messages dominating over Instagram and YouTube content. This may introduce bias and under-represent the influence of less popular platforms on support diffusion. Additionally, while other datasets rely on a keyword-based data collection, our collection of YouTube data is limited to links that were shared in other platforms, which may lead to a less comprehensive sample. Moreover, our study only focuses on four social media platforms and online support shared via URLs, overlooking the potential impact of other platforms and multimedia content. Finally, our analysis of crowdsourcing and crowdfunding URLs only consider the top 100 shared URLs of crowdsourcing and crowdfunding sites, which may have overlooked vulnerable groups who received less attention from the public.

\subsection{Discussion and Future Work}
This study highlights the importance of understanding and leveraging the built-in functionalities and affordances of social media platforms in disseminating different types of online support, thus leading to relevant future work directions. First, despite the high volume of support sites shared on social media platforms, the overall proportion of support URLs is small, accounting for less than 1\% on Twitter and Facebook and less than 3\% on Instagram and YouTube. This fact points out that there may be other important forms of online support other than sharing links, like Facebook groups \cite{carlsen2022ukrainian} and Telegram channels \cite{nazaruk2022subscribe}. Thus, future work may consider including instant messaging apps like Telegram in the analysis of support diffusion. 
Second, our Granger causality analysis focuses on the temporal aspect of sharing support links, implying causal dynamics between social media platforms. However, more advanced statistical frameworks could be used in future research to further analyze the evolution and impact of these activities across various online communities.
Finally, our next research will examine the spread of misinformation across social networks and compare its diffusion, temporal patterns, and cross-platform dynamics with support actions, also looking at the pivotal role of influential users and superspreaders~\cite{nogara2022disinformation, deverna2022superspreaders} in both online support and misinformation activities.


\section{Acknowledgments}
Work supported in part by
DARPA (contract \#HR001121C0169).

\bibliography{aaai22.bib}

\onecolumn
\section{Appendix}
\vspace{1cm}
In Table \ref{web_domain}, we specify the complete list of the web domains used for compiling the list of support sites in this
study. 
In Table \ref{support_site_list}, we show a list of the 51 support sites identified for analysis in this study, grouped into categories.

\begin{table*}[htb]
\centering
\small
\begin{tabular}{| m{15em} | m{30em} |} 
  \hline
  \textbf{Domain} & \textbf{URL} \\
  \hline
  NPR & https://www.npr.org/2022/02/25/1082992947/ukraine- support-help \\
  \hline
  The Washington Post & https://www.washingtonpost.com/world/2022/02/27/how-to- help-ukraine/ \\
  \hline
  CNBC &  https://www.cnbc.com/2022/03/09/heres-a-list-of-top-rated- charities-to-help-the-ukraine-relief-effort.html \\
  \hline
  The Official Website of War in Ukraine &  https://war.ukraine.ua/donate/, https://war.ukraine.ua/support- ukraine/ \\
  \hline
  PayPal’s Support Ukraine Relief page &   https://www.paypal.com/fundraiser/117811171978246391 \\
  \hline
  The Ukrainian Institute London &    https://ukrainianinstitute.org.uk/russias-war-against-ukraine- what-can-you-do-to-support-ukraine-ukrainians/ \\
  \hline
\end{tabular}
\caption{Web domains used for compiling the list of support sites in this
study.}
\label{web_domain}
\end{table*}

\begin{table*}[htb]
\centering
\small
\begin{tabular}{ | m{7em} | m{9em} | m{10cm}|} 
  \hline
  \multicolumn{2}{|c|}{Crowdsourcing Sites} & docs.google.com (Google Docs), forms.gle (Google Forms) \\
  \hline
  \multirow{4}{0cm}{Fundraising Sites} & Crowdfunding Platforms & patreon.com (Patreon), paypal.com (PayPal), gofundme.com (GoFundMe), stripe.com (Stripe), kickstarter.com (Kickstarter), indiegogo.com (Indiegogo), venmo.com (Venmo), ko-fi.com (Ko-fi), fundly.com (Fundly), donately.com (Donately), tipeee.co (Tipeee) \\
  \cline{2-3}
  & 
  Government Fundraising Sites & 
  bank.gov.ua (National Bank of Ukraine), help.gov.ua (Ukraine Government Humanitarian Aid Website), moz.gov.ua (Ministry of Health of Ukraine), u24.gov.ua (UNITED24)
\\ 
  \cline{2-3}
  &
  Local NGOs/Funds & 
  savelife.in.ua (Come Back Alive), voices.org.ua (Voices of Children), razomforukraine.org (Razom Emergency Response), prytulafoundation.org (Serhiy Prytula Charity Foundation), hospitallers.life (The Hospitallers Battalion), supportukrainenow.org (Support Ukraine Now), standforukraine.com (Stand for Ukraine), rsukraine.org (Revived Soldiers Ukraine), redcross.org.ua (Ukrainian Red Cross Society), everybodycan.com.ua (EverybodyCan), lgbtmilitary.org.ua (LGBT Military), novaukraine.org (Nova Ukraine), koloua.com (KOLO Ukraine)
\\ 
  \cline{2-3}
  &
  International NGOs/Funds & war.ukraine.ua (Official Website of Ukraine), crisisrelief.un.org (UN Crisis Relief), unhcr.org (UN Refugee Agency), unicefusa.org (UNICEF), care.org (CARE), doctorswithoutborders.org (Doctors Without Borders), globalgiving.org (GlobalGiving Foundation), rescue.org (International Rescue Committee), icrc.org (International Committee of the Red Cross), projecthope.org (Project Hope), savethechildren.org (Save the Children), wfp.org (World Food Programme), internationalmedicalcorps.org (International Medical Corps), mercycorps.org (Mercy Corps), directrelief.org (Direct Relief), actionagainsthunger.org (Action Against Hunger), helpukraine.center (Help Ukraine Center), unitedhelpukraine.org (United Help Ukraine), airbnb.org (Airbnb.org), sunflowerofpeace.com (Sunflower of Peace Foundation), hias.org (HIAS INC.)
\\ 
  \hline
\end{tabular}
\caption{List of online support sites.}
\label{support_site_list}
\end{table*}

In Table \ref{top10_table}, we provide a list of the top ten support site domains that were shared the most across all four social media platforms.

\begin{table*}[htb]
\centering
\small
\begin{tabular}{ccc} 
  \hline
  \textbf{Domain} & \textbf{Support Site} & \textbf{\# of URLs} \\
  \hline
  patreon.com & Patreon & 69,826 \\
  \hline
  gofundme.com & GoFundMe & 36,631 \\
  \hline
  docs.google.com & Google Docs & 30,460 \\
  \hline
  unhcr.org & UNHCR & 28,330 \\
  \hline
  bank.gov.ua & National Bank of Ukraine & 27,394 \\
  \hline
  war.ukraine.ua & Official Website of War in Ukraine & 25,312 \\
  \hline
  forms.gle & Google Forms & 24,432 \\
  \hline
  paypal.com & PayPal & 22,996 \\
  \hline
  savelife.in.ua & Come Back Alive & 21,687 \\
  \hline
  icrc.org & International Committee of the Red Cross & 10,433 \\
  \hline
\end{tabular}
\caption{A list of the top ten support site domains that were shared the most across all four social media platforms.}
\label{top10_table}
\end{table*}

In Table \ref{granger_table}, we present pairwise Granger causality results between platforms.

\begin{table*}[htb]
\centering
\small
\begin{tabular}{cccm{9cm}ccc} 
  \hline
  \textbf{Source} & \textbf{Destination} & \textbf{\#} & \textbf{Domains} & \textbf{CS} & \textbf{CP} & \textbf{OCS}\\
  \hline
  Twitter & Facebook & 6 & fundly.com{$^{**}$}, rsukraine.org{$^{*}$}, globalgiving.org{$^{**}$}, mercycorps.org{$^{*}$}, novaukraine.org{$^{**}$}, moz.gov.ua{$^{*}$} & 0 & 1 & 5\\
  \hline
  Facebook & Twitter & 6 & docs.google.com{$^{*}$}, rescue.org{$^{**}$}, unhcr.org{$^{*}$}, redcross.org.ua{$^{**}$}, directrelief.org{$^{*}$}, unitedhelpukraine.org{$^{*}$} & 1 & 0 & 5\\
  \hline
  Twitter & Instagram & 7 & docs.google.com{$^{*}$}, forms.gle{$^{***}$}, patreon.com{$^{*}$}, globalgiving.org{$^{*}$}, mercycorps.org{$^{*}$}, moz.gov.ua{$^{**}$}, doctorswithoutborders.org{$^{**}$} & 2 & 1 & 4\\
  \hline
  Instagram & Twitter & 6 & voices.org.ua{$^{**}$}, care.org{$^{***}$}, icrc.org{$^{***}$}, directrelief.org{$^{*}$}, airbnb.org{$^{*}$}, sunflowerofpeace.com{$^{*}$} & 0 & 0 & 6\\
  \hline
  Twitter & YouTube & 14 & docs.google.com{$^{*}$}, patreon.com{$^{*}$}{$^{*}$}, venmo.com{$^{*}$}, tipeee.co{$^{*}$}, voices.org.ua{$^{***}$}, rsukraine.org{$^{***}$}, globalgiving.org{$^{***}$}, rescue.org{$^{*}$}, prytulafoundation.org{$^{*}$}, crisisrelief.un.org{$^{***}$}, hospitallers.life{$^{***}$}, novaukraine.org{$^{*}$}, koloua.com{$^{*}$}, standforukraine.com{$^{***}$} & 1 & 3 & 10\\
  \hline
  YouTube & Twitter & 8 & paypal.com{$^{**}$}, ko-fi.com{$^{*}$}, bank.gov.ua{$^{**}$}, savethechildren.org{$^{*}$}, doctorswithoutborders.org{$^{**}$}, care.org{$^{*}$}, unitedhelpukraine.org{$^{*}$}, moz.gov.ua{$^{***}$} & 0 & 2 & 6\\
  \hline
  Facebook & Instagram & 8 & kickstarter.com{$^{**}$}, venmo.com{$^{**}$}, ko-fi.com{$^{*}$}, helpukraine.center{$^{***}$}, internationalmedicalcorps.org{$^{*}$}, redcross.org.ua{$^{***}$}, directrelief.org{$^{*}$}, hias.org{$^{***}$} & 0 & 3 & 5\\
  \hline
  Instagram & Facebook & 6 & voices.org.ua{$^{***}$}, rsukraine.org{$^{***}$}, care.org{$^{*}$}, moz.gov.ua{$^{**}$}, actionagainsthunger.org{$^{*}$}, sunflowerofpeace.com{$^{**}$} & 0 & 0 & 6\\
  \hline
  Facebook & YouTube & 8 & gofundme.com{$^{*}$}, bank.gov.ua{$^{*}$}, helpukraine.center{$^{*}$}, unhcr.org{$^{*}$}, standforukraine.com{$^{**}$}, rsukraine.org{$^{***}$}, directrelief.org{$^{*}$}, koloua.com{$^{*}$} & 0 & 1 & 7\\
  \hline
  YouTube & Facebook & 8 & paypal.com{$^{*}$}, venmo.com{$^{*}$}, globalgiving.org{$^{***}$}, savethechildren.org{$^{*}$}, unicefusa.org{$^{*}$}, internationalmedicalcorps.org{$^{**}$}, hospitallers.life{$^{**}$}, supportukrainenow.org{$^{**}$} & 0 & 2 & 6\\
  \hline
  Instagram & YouTube & 12 & savelife.in.ua{$^{***}$}, everybodycan.com.ua{$^{***}$}, supportukrainenow.org{$^{*}$}, rsukraine.org{$^{***}$}, care.org{$^{*}$}, globalgiving.org{$^{**}$}, icrc.org{$^{**}$}, projecthope.org{$^{***}$}, doctorswithoutborders.org{$^{*}$}, redcross.org.ua{$^{*}$}, novaukraine.org{$^{***}$}, moz.gov.ua{$^{***}$} & 0 & 0 & 12\\
  \hline
  YouTube & Instagram & 5 & bank.gov.ua{$^{*}$}, rescue.org{$^{***}$}, wfp.org{$^{**}$}, unhcr.org{$^{***}$}, prytulafoundation.org{$^{***}$} & 0 & 0 & 5\\
  \hline
\end{tabular}
\caption{Pairwise Granger causality results between platforms. The maximum lag parameter is set to 3 and the significance level to \emph{p} $<$ .05 (Note: {$^{*}$} \emph{p} $<$ .05, {$^{**}$} \emph{p} $<$ .01, {$^{***}$} \emph{p} $<$ .001). \textbf{\#} = Total number of support sites that the source platform Ganger-causes the destination platform, \textbf{CS} = Crowdsourcing Sites (Google Docs and Forms), \textbf{CP} = Crowdfunding Platforms, \textbf{OCS} = Other Crowdfunding Sites (Government, Local and International NGOs/Funds). \# = \textbf{CS} + \textbf{CP} + \textbf{OCS}.}
\label{granger_table}
\end{table*}

\end{document}